\documentstyle[12pt,epsfig]{article}
\def\be{\begin{equation}}
\def\ee{\end{equation}}
\def\bea{\begin{eqnarray}}
\def\eea{\end{eqnarray}}

\begin{document}
\def\singlespacing{\baselineskip=12pt}
\def\doublespacing{\baselineskip=20pt}
\doublespacing
\pagestyle{empty}

\noindent
August 30th, 1996 \hfill {\bf MC-TH-96/27}

\vspace{2cm}

\begin{center} 
\begin{large}

{\bf PION POLARIZABILITY AND HADRONIC TAU DECAYS}

\bigskip

{\em V. Kartvelishvili\/${}^{a,b}$, M. Margvelashvili\/${}^{a}$  \\
 and G. Shaw\/${}^{b}$}

\end{large}

\bigskip

${}^{a}$~High Energy Physics Institute, Tbilisi State University,\\
Tbilisi, GE-380086, Republic of Georgia.

\bigskip

${}^{b}$~Department of Physics and Astronomy, Schuster Laboratory,\\
University of Manchester, Manchester M13 9PL, U.K.

\end{center}

\bigskip

\bigskip

\begin{center}
{\bf ABSTRACT}
\end{center}

\noindent
Recent experimental  data on $\tau$ decays are used  
to reconstruct the difference in hadronic spectral densities with vector
and axial-vector quantum numbers. The saturation of Das-Mathur-Okubo and 
Weinberg sum rules is studied. Two methods of improving convergence
and decreasing errors are applied, and good agreement with the
predictions of current algebra and chiral perturbation theory is
observed. The resulting value of the pion polarisability is 
$\alpha_E = (2.64 \pm 0.36) \, 10^{-4} \; {{\rm fm}}^3$.

\vspace{5cm}

{\em Paper presented at the \lq\lq QCD Euroconference'', Montpellier,
July 4-12th, 1996.}

\newpage
\pagestyle{plain}
\pagenumbering{arabic}

\section{Introduction}

The electric polarisability of the charged pion $\alpha_E$ can be inferred
from the amplitude for low energy Compton scattering 
$ \gamma + \pi^+ \rightarrow \gamma + \pi^+ $.
 This amplitude cannot be measured
 at low energies directly, but can be determined from measurements on
related processes like $\pi N \rightarrow \pi N \gamma$,  $ \gamma N 
\rightarrow \gamma N \pi$  and $ \gamma \gamma \rightarrow \pi \pi$.
The measured values for $\alpha_E$, 
in units of $10^{-4}\; {{\rm fm}}^3$, 
are $6.8 \pm 1.4 \pm 1.2$ \cite{Antipov}, 
$ 20 \pm 12$ \cite{Aibergenov} and $2.2 \pm 1.6$ \cite{Babusci}, respectively. 

Alternatively, the polarisability can be predicted theoretically by relating 
it to other
quantities which are better known experimentally. In chiral perturbation
theory, it can be shown \cite{Holstein} that the pion polarisability
is given by
\be
\alpha_E = \frac{ \alpha  \, F_A}{m_{\pi} F_{\pi}^2},
\label{alpha}
\ee
where $F_{\pi}$ is the pion decay 
constant and $F_A$ is the axial structure-dependent form factor  for radiative 
charged pion decay $\pi \rightarrow e \nu \gamma$ \cite{Bryman}. 
The latter is often re-expressed in the form 
\be
F_A \equiv \gamma\,{F_V}\, ,
\ee
because the ratio $\gamma$ can be measured
 in radiative pion decay experiments more accurately than $F_A$ itself, 
while the corresponding vector form factor $F_V$  is determined to be 
\cite{PDG}\footnote{Our definitions of $F_V$ and $F_A$ differ from
those used by the Particle Data Group \cite{PDG} by a factor of two.}
$F_V  = 0.0131 \pm 0.0003$
by using  the conserved
vector current (CVC) hypothesis to relate $\pi \rightarrow e \nu \gamma$
and $\pi^0 \rightarrow  \gamma\gamma$ decays. 
$\gamma$ has been measured in three $\pi \rightarrow e \nu \gamma$ 
experiments, giving the values: $0.25 \pm 0.12$ \cite{lampf}; 
$0.52 \pm 0.06$ \cite{sin}; and $0.41 \pm 0.23$ \cite{istra}. The weighted 
average $ \gamma = 0.46 \pm 0.06$
can be combined with the above equations to give
\be
\alpha_E = (2.80 \pm 0.36) \, 10^{-4} \; {{\rm fm}}^3 \; .
\label{chiral}
\ee

This result is often referred to as the chiral theory prediction for
the pion polarisability \cite{Holstein}. However $\alpha_E$, or 
equivalently $F_A$, can be 
determined in
other ways. In particular, the latter occurs in the  Das-Mathur-Okubo
(DMO) sum rule \cite{Das}                     
\be
I = F_{\pi}^2  \frac{\langle r_{\pi}^2 \rangle }{3} - {F_A}\,, 
\label{DMO}
\ee
where
\be
I\equiv\int \frac{{\rm d}s}{s} \rho_{V-A}(s) \,,
\label{i}
\ee
with $\rho_{V-A}(s) = \rho_V(s) - \rho_A(s)$ being the difference 
in the spectral functions of the vector and axial-vector
isovector current correlators, while
$\langle r_{\pi}^2\rangle $
is the pion mean-square charge radius. Using its
standard value $\langle r_{\pi}^2\rangle = 0.439 \pm 0.008 \; {{\rm fm}}^2$
 \cite{Amendolia} and eqs. (\ref{alpha}), (\ref{chiral}) one gets:
\be\label{iexp}
I_{DMO}=(26.6 \pm 1.0)\cdot 10^{-3}
\ee

Alternatively, if the integral $I$ is known, eq. (\ref{DMO}) can be rewritten
in the form of a prediction for the polarisability:
\be 
\alpha_E  = \frac{\alpha}{m_{\pi}} \biggl( 
\frac{\langle r_{\pi}^2 \rangle }{3}
- \frac{I}{F_{\pi}^2} \biggr).
\label{DMOpredict}
\ee 

Recent attempts to analyse this relation have resulted in some contradiction
with the chiral prediction. 
Lavelle et al. \cite{Lavelle} 
use related QCD sum rules to estimate the integral $I$ and obtain
$\alpha_E = (5.60 \pm 0.50) \, 10^{-4} \; {{\rm fm}}^3$.
Benmerrouche et. al. \cite{Bennerrouche} apply certain sum rule
inequalities to obtain a lower bound on the polarisability 
(\ref{DMOpredict}) as a function of 
${\langle r_{\pi}^2 \rangle }$. Their analysis also
tends to prefer larger $\alpha_E$ and/or smaller 
${\langle r_{\pi}^2\rangle }$ values.

In the following we use available experimental data to reconstruct the 
hadronic spectral function $\rho_{V-A}(s)$, in order to calculate
the integral 
\be\label{i0}
I_0(s_0) \equiv  \int_{4m_{\pi}^2}^{s_0} 
\frac{{\rm d}s}{s} \rho_{V-A}(s) \; .       
\ee
for $s_0\simeq M_{\tau}^2$, test the saturation of the DMO sum rule
(\ref{DMO}) and its compatibility with the chiral prediction (\ref{chiral}).
We also test the saturation of the first Weinberg sum rule \cite{Weinberg}:
\bea
W_1(s_0)  \equiv  \int_{4m_{\pi}^2}^{s_0} 
{\rm d}s \, \rho_{V-A}(s)\;;\;\;\;\;\;\;\;\;\;\; 
W_1(s_0)\;\bigl|_{s_0\to\infty}  = F_{\pi}^2 
\label{w1}
\eea
and use the latter to improve convergence and 
obtain a more accurate estimate for the integral $I$:
\begin{eqnarray}\label{i1}
I_1(s_0) & = & \int_{4m_{\pi}^2}^{s_0} \frac{{\rm d}s}{s} 
\rho_{V-A}(s) \nonumber \\
& + & {{\beta}\over{s_0}} 
\left[ F_{\pi}^{2} - \int_{4m_{\pi}^2}^{s_0} {\rm d}s \, \rho_{V-A}(s) 
\right] 
\end{eqnarray}
Here the parameter $\beta$ is arbitrary and can be chosen to minimize the
estimated error in $I_1$ \cite{markar}. 

Yet another way of reducing the uncertainty in our estimate of $I$ is to
use the Laplace-transformed version of the DMO sum rule \cite{marg}: 
\be
I_2(M^2) = F_{\pi}^2  \frac{\langle r_{\pi}^2 \rangle}{3} - {F_A} 
\ee
with $M^2$ being the Borel parameter in the integral
\begin{eqnarray}
I_2(M^2) &\equiv& \int \frac{{\rm d}s}{s} \exp{\left( \frac{-s}{M^2} \right)} \, 
\rho_{V-A}(s) + \frac{F_{\pi}^2}{M^2} \nonumber \\ 
& - & \frac{C_6 \langle O^6\rangle}{6 M^6} - 
\frac{C_8 \langle O^8\rangle }{24 M^8} + \ldots \;.
\label{i2}
\end{eqnarray}
Here $ C_6 \langle O^6\rangle $ and $ C_8 \langle O^8\rangle $
are the four-quark vacuum condensates of dimension 6 and 8, whose values
could be estimated theoretically or taken from previous analyses
\cite{markar,kar}.

All the three integrals (\ref{i0}), (\ref{i1}) and (\ref{i2}) obviously reduce
to (\ref{i}) as $s_0, M^2 \to \infty$.

\section{Evaluation of the spectral densities}

Recently ALEPH  published a comprehensive and consistent
set of $\tau$ branching fractions \cite{aleph}, where in many cases 
the errors are
smaller than previous world averages. We have used these values to 
normalize the contributions of specific hadronic final states, while 
various available experimental data have been used to determine the 
shapes of these contributions. Unless stated otherwise, 
each shape was fitted with a single
relativistic Breit-Wigner distribution with appropriately chosen threshold
behaviour.

\subsection{ Vector current contributions.}

A recent comparative study {\cite{eidel}}
of corresponding final states in $\tau$ decays and
$e^+e^-$ annihilation
has found no significant violation of CVC or isospin symmetry. 
In order to determine the shapes of the hadronic spectra, we have used
mostly $\tau$ decay data, complemented by $e^+e^-$ data in some
cases.

{{ $\pi^-\pi^0$ :}} ${{\rm BR}}=25.30\pm0.20\%$ \cite{aleph}, 
and the $s$-dependence was
described by the three interfering resonances $\rho(770)$, $\rho(1450)$ 
and $\rho(1700)$, with the parameters taken from 
\cite{PDG} and \cite{pi2}.

{{ $3\pi^{\pm}\pi^0$ :}} ${{\rm BR}}=4.50\pm0.12\%$, including $\pi^-\omega$ 
final state \cite{aleph}. The shape was determined by 
fitting the spectrum measured by ARGUS  \cite{pi4a}.

{{ $\pi^{-}3\pi^0$ :}} ${{\rm BR}}=1.17\pm0.14\%$ \cite{aleph}. 
The $s$-dependence
is related to that of the reaction $e^+e^-\to 2\pi^+2\pi^-$. We have fitted
the latter measured by OLYA and DM2  \cite{pi4b}.

{{ $6\pi$ :}} various charge contributions give the overall 
${{\rm BR}}=0.13\pm0.06\%$ \cite{aleph}, fairly close to CVC expectations
\cite{eidel}.

{{ $\pi^-\pi^0\eta$ :}} ${{\rm BR}}=0.17\pm0.03\%$ \cite{pi0}. 
The $s$-dependence
was determined by fitting  the distribution measured by CLEO 
\cite{pi0}.

{{ $K^-K^0$ :}} ${{\rm BR}}=0.26\pm0.09\%$ \cite{aleph}. 
Again, the fit of the CLEO  measurement \cite{cleok} was performed.

\subsection{ Axial current contributions.}

The final states with odd number of pions contribute to the axial-vector
current. 
Here, $\tau$ decay is the only source of precise information. 

{{ $\pi^-$ :}} ${{\rm BR}}=11.06\pm0.18\%$ \cite{aleph}. The single pion
contribution has a trivial $s$-dependence and hence is
explicitly taken into account in theoretical formulae. The quoted
branching ratio corresponds to $F_{\pi}=93.2$ MeV. 

{{ $3\pi^{\pm}$ and $\pi^-2\pi^0$ :}} 
${{\rm BR}}=8.90\pm0.20\%$ and ${{\rm BR}}=9.21\pm0.17\%$, 
respectively \cite{aleph}. Theoretical models \cite{pi3th} 
assume that these two modes are identical in both shape and normalization. 
The $s$-dependence has been analyzed in \cite{opal},
where the parameters of two theoretical models describing this decay
have been determined. We have used the average of these two distributions,
with their difference taken as an estimate of the shape uncertainty.

{{ $3\pi^{\pm}2\pi^0$ :}} ${{\rm BR}}=0.50\pm0.09\%$, 
including  $\pi^-\pi^0\omega$ 
final state \cite{aleph}. The shape was fitted using the
CLEO measurement \cite{pi5a}.

{{ $5\pi^{\pm}$ and $\pi^-4\pi^0$ :}} 
${{\rm BR}}=0.08\pm0.02\%$ and $BR=0.11\pm0.10\%$, 
respectively \cite{aleph}. We have assumed that these two terms have the same
$s$-dependence measured in \cite{pi5b}.

\subsection{ $K{\overline K} \pi$ modes.}

$K{\overline K} \pi$ modes can
contribute to both vector and axial-vector currents, and various theoretical
models cover the widest possible range of predictions \cite{kkpith}.

According to \cite{aleph}, all three $K{\overline K} \pi$ modes 
(${\overline K}^0K^0\pi^-, K^-K^0\pi^0$ and
$K^-K^+\pi^-$) add up to BR$({\overline K} K \pi)=0.56\pm0.18\%$, in 
agreement with other measurements (see \cite{cleok}). 
The measured $s$-dependence  suggests that these final 
states are dominated by $K^*K$ decays \cite{cleok}.
We have fitted the latter, taking into
account the fact that due to parity constraints, vector and axial-vector  
$K^*K$ terms have different threshold behaviour. A parameter
$\xi$ was defined as the portion of $K{\overline K} \pi$ final state with
axial-vector quantum numbers, so that
\begin{eqnarray}
{\rm BR}({\overline K} K\pi)_{V}&=&(1-\xi)\;{\rm BR}({\overline K}K\pi)
\nonumber\\
{\rm BR}({\overline K} K\pi)_{A}&=&\xi\;{\rm BR}({\overline K} K\pi)\,.
\end{eqnarray}

\section{Results and conclusions}

% REFERENCES TO EQUATIONS HAVE TO BE PUT IN BY HAND IN CAPTIONS BELOW

\begin{figure}[htb]
\begin{center}
\epsfig{file=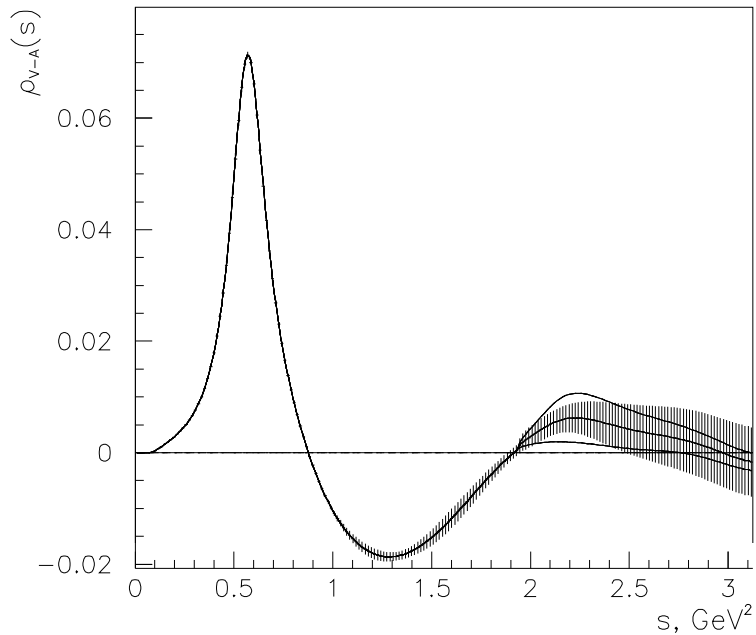,width=10cm,clip=}
\end{center}
\vspace{1cm}
\caption{\tenrm
Difference of vector and axial-vector hadronic spectral densities.
In figs.1-5: the three curves  correspond
to $\xi=0$, $0.5$ and 1 from top to bottom;
the errors originating from the shape variation and those coming
from the errors in the branching fractions are roughly equal and have
been added in quadrature to form the error bars, shown only for $\xi=0.5$.}
\label{fig1}
\end{figure}

\begin{figure}[p]
\begin{center}
\epsfig{file=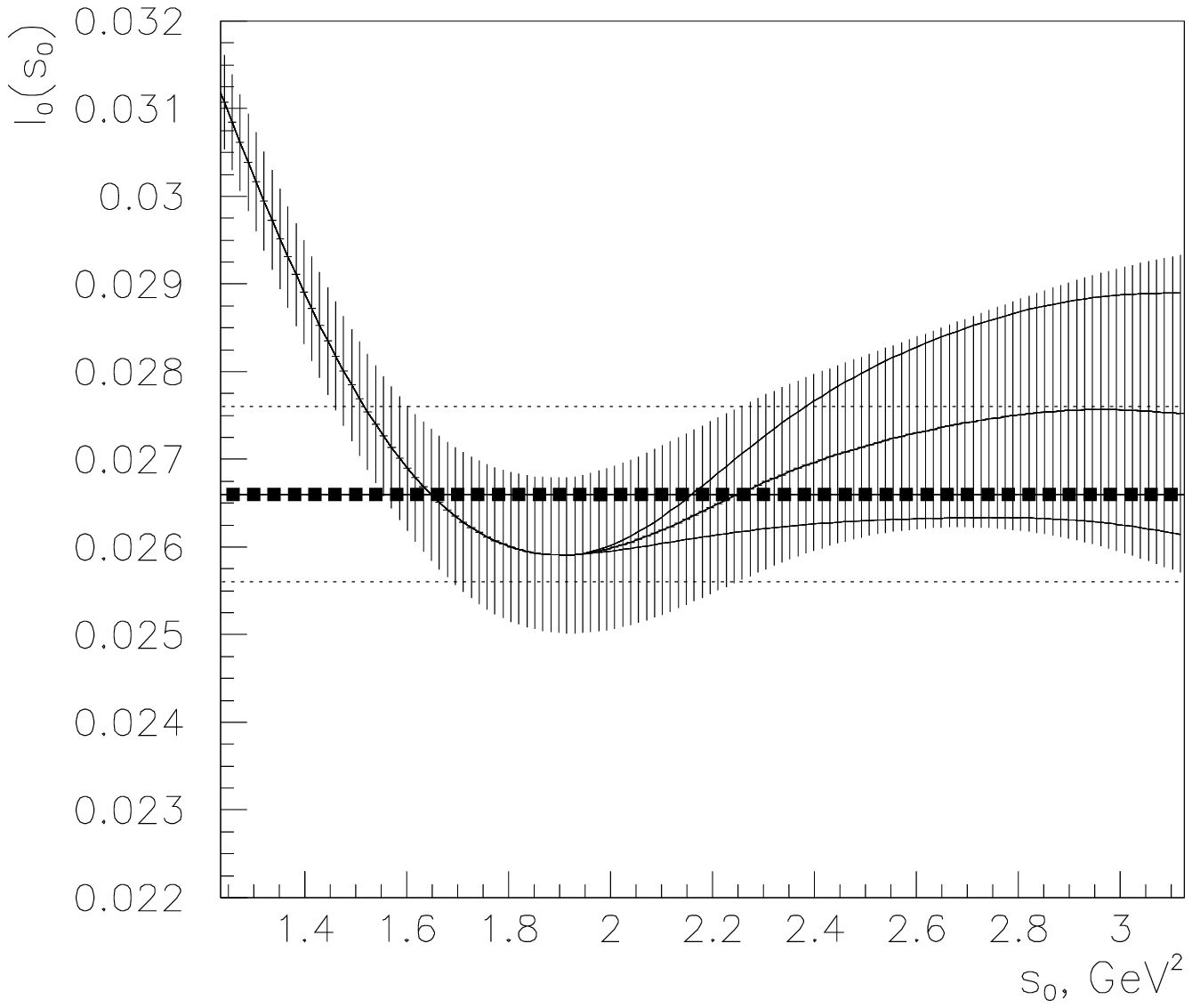,width=10cm,clip=}
\end{center}
\vspace{1cm}
\caption{\tenrm
Saturation of the DMO sum rule integral (8). 
The thick dashed line is the chiral prediction for the asymptotic value
(6) and the dotted lines show its errors.}
\label{fig2}
%\end{figure}
%
%\begin{figure}[htb]
\begin{center}
\epsfig{file=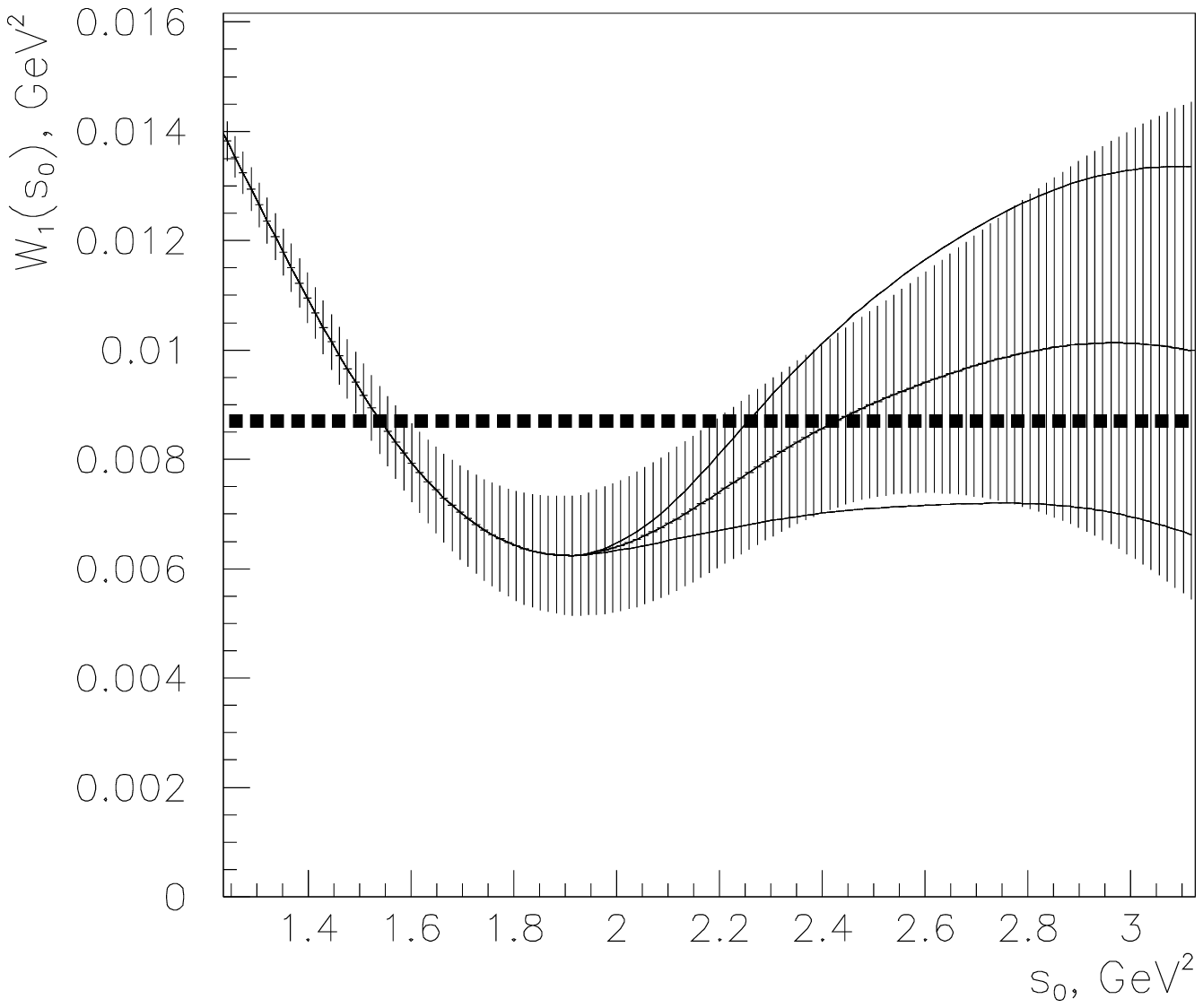,width=10cm,clip=}
\end{center}
\vspace{1cm}
\caption{\tenrm
Saturation of the first Weinberg sum rule (9).
 The dashed line shows the expected asymptotic value $F_{\pi}^2$.}
\label{fig3}
\end{figure}

The resulting spectral function is shown in Fig.1.
The results of its integration according to (\ref{i0})
are presented in fig.2 as a function of the upper bound $s_0$. 
One can see that as $s_0$
increases, $I_0$ converges towards an asymptotic value which we
estimate to be{\footnote{In the following, the first error
corresponds to the quadratic sum of the errors in the branching ratios
and the assumed shapes, while the second one arises from to the variation
of $\xi$ in the interval $0.5\pm 0.5$.}
\be\label{i0m}
I_0 \equiv I_0(\infty) = ( 27.5 \pm 1.4 \pm 1.2 ) \cdot 10^{-3},
\ee
in good agreement with the chiral value (\ref{iexp}).

The saturation of the Weinberg sum rule (\ref{w1}) is shown in fig.3.
One sees that the expected value $F_{\pi}^2$ is well within the errors,
and $\xi \simeq 0.25\div0.3$ seems to be preferred. No
significant deviation from this sum rule is expected theoretically
\cite{Floratos}, so we use (\ref{i1}) to calculate our second
estimate of the integral $I$. The results of this
integration are presented in fig.4, with the asymptotic value  
\be\label{i1m}
I_1 \equiv I_1(\infty) = ( 27.0 \pm 0.5 \pm 0.1 ) \cdot 10^{-3},
\ee
corresponding to $\beta\approx 1.18$.
One sees that the convergence has improved, the errors 
are indeed much smaller, and  the $\xi$-dependence is very weak.

\begin{figure}[p]
\begin{center}
\epsfig{file=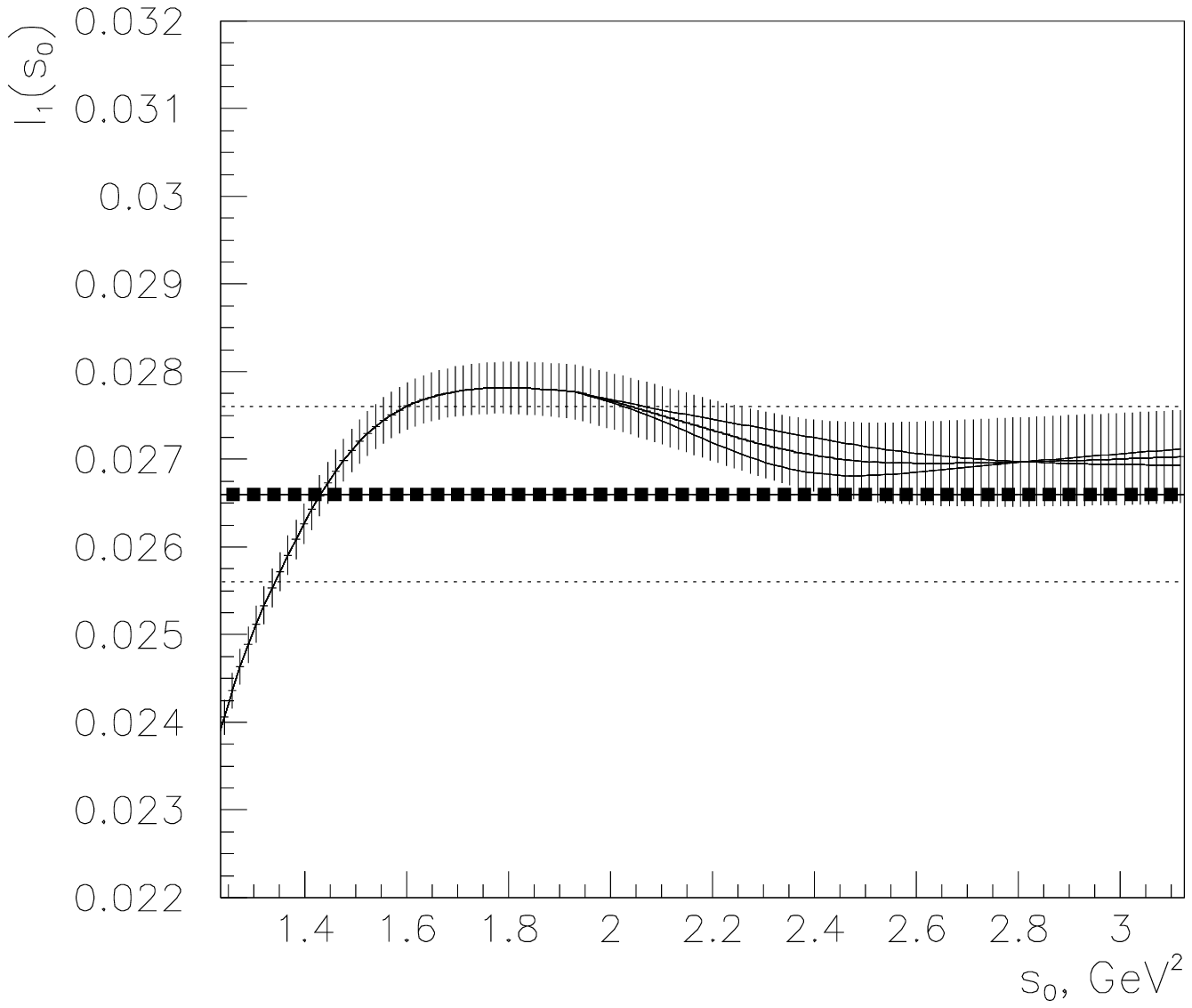,width=10cm,clip=}
\end{center}
\vspace{1cm}
\caption{\tenrm
Saturation of the modified DMO sum rule integral (10). 
The chiral prediction is also shown as in fig.2.}
\label{fig4}
%\end{figure}
%
%\begin{figure}[htb]
\begin{center}
\epsfig{file=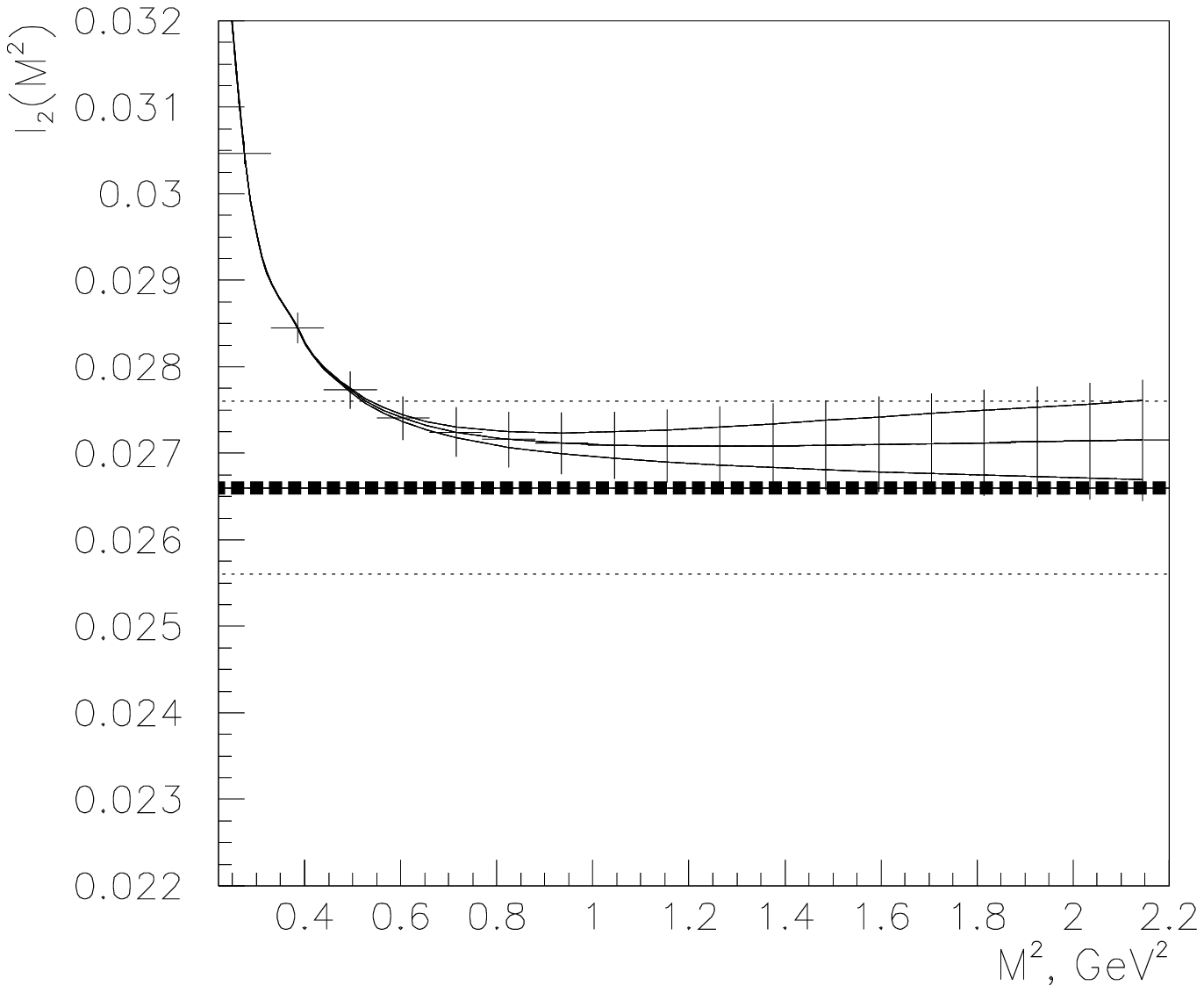,width=10cm,clip=}
\end{center}
\vspace{1cm}
\caption{\tenrm
The Laplace-transformed sum rule (12) as a function of 
the Borel parameter $M^2$, compared
to the chiral prediction.}
\label{fig5}
\end{figure}

Now we use (\ref{i}) to obtain our third estimate of the spectral 
integral. The integration results are plotted against the 
 Borel parameter $M^2$ in fig.5, assuming standard values for dimension
6 and 8 condensates. 
The results are independent
of $M^2$ for $M^2 > 1 \, GeV^2$,  indicating that  higher order terms
are negligible in this region, and giving
\be\label{i2m}
I_2\equiv I_2(\infty) = ( 27.2 \pm 0.4 \pm 0.2 \pm 0.3) \, 10^{-3} \; ,
\ee
where the last error reflects the sensitivity of (\ref{i}) to the variation
of the condensate values.

One sees that these three numbers (\ref{i0m}) -- (\ref{i2m}) are in 
good agreement with each other and with the chiral prediction (\ref{iexp}).
Substitution of our most precise result (\ref{i1m}) into (\ref{DMOpredict})
yields for the standard value of the pion charge radius quoted above:
\be\label{alem} 
\alpha_E = ( 2.64 \pm 0.36 ) \, 10^{-4} \; {{\rm fm}}^3, 
\ee
in good  agreement
with (\ref{chiral}) and the smallest of the measured values, \cite{Babusci}. 
Note that by substituting a  larger value 
$\langle r_{\pi}^2\rangle = 0.463 \pm 0.006 \; {{\rm fm}}^2$ \cite{gesh},
one obtains 
$\alpha_E = (3.44 \pm 0.30) \, 10^{-4} \; {{\rm fm}}^3$, 
about two standard deviations higher than 
(\ref{chiral}).

In conclusion, we have used recent precise  data 
to reconstruct the difference in vector and axial-vector hadronic
spectral densities and to study the saturation of Das-Mathur-Okubo 
and the first Weinberg sum rules. Two methods of improving convergence
and decreasing the errors have been used.
Within the present level of accuracy, we have found perfect consistence
between $\tau$ decay data, chiral and QCD sum rules, the standard value
of $\langle r_{\pi}^2\rangle$, the average value of $\gamma$ and the chiral
prediction for $\alpha_E$.

Helpful discussions and correspondence with R. Alemany, R. Barlow, M.Lavelle
 and P. Poffenberger are gratefully acknowledged.


\newpage

\begin{thebibliography}{99}

\bibitem{Antipov} Y.M.Antipov et al. {\em Z. Phys.} {\bf C26} (1985) 495.

\bibitem{Aibergenov} T.A.Aibergenov et al.  {\em Czech J. Phys.} {\bf B36} 
(1986) 948.

\bibitem{Babusci}D.Babusci et al. {\em Phys.Lett.} {\bf B277}
(1992) 158.

\bibitem{Holstein}   M.V.Terent'ev, {\em Sov. J. Nucl. Phys.} {\bf 16} (1972)
162; B.R.Holstein, {\em Comm. Nucl. Part. Phys.} {\bf 19} (1990) 221.

\bibitem{Bryman} D.A.Bryman et al.  {\em Phys. Rep.} {\bf 88}
(1982) 151.

\bibitem{PDG} Particle Data Group {\em Phys. Rev.} {\bf D50} (1994) 1172. See
especially pp. 1447,1448.

\bibitem{lampf} L.E.Piilonen et al.,  {\em Phys.Rev.Lett.} {\bf 57}
(1986) 1402.

\bibitem{sin} A.Bay et al. {\em Phys. Lett.} {\bf B174}
(1986) 445.

\bibitem{istra} V.N.Bolotov et al., {\em Phys.Lett.} {\bf B243}
(1990) 308.

\bibitem{Das} T.Das, V.Mathur and S.Okubo, {\em Phys. Rev. Lett.} {\bf 19}
(1967) 859; A.I.Vainshtein, {\em JETP Lett.} 
{\bf 6} (1967) 815; {\bf 7} (1967) 81 (E). 

\bibitem{Amendolia} S.R.Amendolia et al., {\em Phys. Lett.} {\bf B178}
(1986) 244.                                  
 
\bibitem{Lavelle} M.O.Lavelle, N.F.Nasrallah and K.Schilcher  
{\em Phys. Lett.} {\bf B335} (1994) 211.

\bibitem{Bennerrouche} M.Benmerrouche, G.Orlandini and T.G.Steele 
{\em Phys. Lett.} {\bf B366} (1996) 354.

\bibitem{Weinberg} S.Weinberg,  {\em Phys. Rev. Lett.} {\bf 18} (1967) 507.

\bibitem{markar} V.Kartvelishvili and M.Margvelashvili, {\em Z. Phys.} 
{\bf C55} (1992) 83.                                  

\bibitem{Floratos} E.Floratos, S.Narison and E.de Rafael, {\em Nucl. Phys.}
{\bf B155} (1979) 115; R.D.Peccei and J.Sol\`a, {\em Nucl. Phys.}
{\bf B281} (1987) 1.           

\bibitem{marg} M.Margvelashvili, {\em Phys.Lett.} {\bf B188}
(1988) 763.                               

\bibitem{kar} C.A.Dominguez and J.Sol\`a,  {\em Z. Phys.} 
{\bf C40} (1988) 63; V.Kartvelishvili, {\em Phys. Lett.} 
{\bf B287} (1992) 159.

\bibitem{aleph} D.Buskulic et al., {\em Z. Phys.} 
{\bf C70} (1996) 579.                                   

\bibitem{eidel} S.I.Eidelman and V.N.Ivanchenko, {\em Nucl. Phys.
 (Proc. Suppl.)} {\bf B40} (1995) 131.

\bibitem{pi2} D.Bisello et al., {\em Phys. Lett.} {\bf B220}
(1989) 321.                                   

\bibitem{pi4a} H.Albrecht et al., {\em Phys.Lett.} {\bf B185}
(1987) 223.                                  

\bibitem{pi4b} L.M.Kurdadze et al., {\em JETP Lett.} 
{\bf 47} (1988) 512; D.Bisello et al., LAL 91-64, Orsay, 1991.

\bibitem{pi0} M.Artuso et al., {\em Phys.Rev.Lett.} {\bf 69} 
(1992) 3278.  

\bibitem{cleok} T.E.Coan et al., {\em Phys. Rev.} 
{\bf D53} (1996) 6037.                 

\bibitem{pi3th} N.Isgur et al.,  {\em Phys. Rev.} 
{\bf D39} (1989) 1357; J.H.K\"uhn and A.Santamaria,  {\em Z. Phys.} 
{\bf C48} (1990) 445.                                  
                                  

\bibitem{opal} R.Akers et al., CERN-PPE/95-022,
Geneva, 1995.                              

\bibitem{pi5a} D.Bortoletto et al., {\em Phys.Rev.Lett.} {\bf 71} 
(1993) 1791.                               

\bibitem{pi5b} D.Buskulic et al., {\em Phys.Lett.} {\bf B349}
(1987) 585.                                

\bibitem{kkpith} J.J.Gomes-Cadenas et al., {\em Phys.Rev.} {\bf D42} 
(1990) 3093; R.Decker et al.,  {\em Z.Phys.} 
{\bf C58} (1993) 445; M.Finkemeier and E.Mirkes,  {\em Z.Phys.} 
{\bf C69} (1996) 243.                                                                                                    

\bibitem{gesh} B.V.Geshkenbein, {\em Z.Phys.} 
{\bf C45} (1989) 351.                                   

\end{thebibliography}
\end{document}